# Influence of Local Correlations on the "Homogeneous Insulator−Superconductor" Transition in the Domain Boundaries of the Charge-Order Phase of a 2D System of a Mixed Valence

V. V. Konev[a],*, V. A. Ulitko[a], D. N. Yasinskaya[a], Yu. D. Panov[a], and A. S. Moskvin[a]

[a] *Ural Federal University, Yekaterinburg, 620002 Russia*
*e-mail: vitaliy.konev@urfu.ru*

**Abstract**—It is demonstrated in the (pseudo)spin $S = 1$ formalism that the structure of antiphase domain boundaries in the phase of charge ordering of a mixed-valence system of the $Cu^{1+, 2+, 3+}$ "triplet" type in cuprates on a two-dimensional square lattice depends to a considerable extent on on-site correlation parameter $U$. The results of computer modeling on large square lattices illustrate the change in the boundary structure (from a homogeneous monovalent nonconducting structure of the $Cu^{2+}$ type to a filamentary superconducting one) induced by a relatively small variation of positive $U$ values.

## 1. INTRODUCTION

The interest in systems with spin $S = 1$ is fuelled by the studies both into high-anisotropy magnetic materials based on $Ni^{2+}$ (e.g., $[Ni(HF_2)(3\text{-}Clpy)_1]BF_1]$ and $NiCl_2 4SC(NH_2)_2$) and into the so-called pseudospin systems of the semi-hard-core boson type with restrictions as to the occupation of lattice sites $n = 0, 1, 2$ or systems of ions with a mixed valence of the "triplet" type ($Cu^{1+, 2+, 3+}$ in cuprates $La_{2-x}Sr_xCuO_4$ or $Bi^{3+, 4+, 5+}$ in bismuthates [1]). In all cases, the phase diagram of spin or pseudospin systems with $S = 1$ is much more complex than that of similar systems with quantum (pseudo)spin $S = 1/2$. The primary reason for this is the emergence of entirely new terms in the Hamiltonian (of the single-ion anisotropy and biquadratic interaction type) and fundamentally new phases such as a quantum paramagnetic or spin-nematic phases.

Depending on the parameters of local and intersite charge–charge correlations, one- and two-particle transfer integrals, and the total charge, the ground state of such systems may correspond to charge ordering, various types of superconducting ordering, composite phases of the supersolid type with coexisting superconductivity and charge ordering, or to a quantum paramagnetic phase, which is specific to these systems. The emergence of various metastable heterogeneous states with a well-developed domain structure and topological defects of the vortex or skyrmion type is typical of such systems [2–4].

In this study, the pseudospin formalism is used to analyze a simple system of $Cu^{1+, 2+, 3+}$ charge triplets in a model cuprate. It is demonstrated that the structure of antiphase domain boundaries in the charge-order phase of this system depends to a considerable extent on on-site correlation parameter $U$ and changes from a homogeneous monovalent nonconducting structure of the $Cu^{2+}$ type to a filamentary superconducting one under a relatively small variation of positive $U$ values.

## 2. MODEL CUPRATE: PSEUDOSPIN $S = 1$ FORMALISM

The model cuprate is a 2D system of Cu sites in the $CuO_2$ cuprate plane with three different possible valence charge states: $Cu^{1+, 2+, 3+}$. This charge triplet is associated with three pseudospin $S = 1$ states in the following way: $Cu^{1+} \rightarrow M_S = -1$, $Cu^{2+} \rightarrow M_S = 0$, $Cu^{3+} \rightarrow M_S = +1$. In further study, we use well-known methods for characterization of spin systems.

The pin algebra of systems with $S = 1$ ($M_S = 0, \pm 1$) includes eight independent nontrivial (three dipole and five quadrupole) operators:

$$S_z;\ S_\pm = \mp \frac{1}{\sqrt{2}}(S_x \pm iS_y);\ S_z^2; \\ T_\pm = \{S_z, S_\pm\} \equiv S_z S_\pm + S_\pm S_z;\ S_\pm^2. \quad (1)$$

Raising and lowering operators $S_\pm$ and $T_\pm$ alter the pseudospin projection by $\pm 1$, but in different ways: $\langle 0|S_\pm|\mp 1\rangle = \langle \pm 1|S_\pm|0\rangle = \mp 1$, $\langle 0|T_\pm|\mp 1\rangle = -\langle \pm 1|T_\pm|0\rangle = +1$. Raising and lowering operators $S_\pm^2$ characterize

the $|-1\rangle \leftrightarrow |+1\rangle$ transitions; i.e., they "produce" a hole ($S_+^2$) or electron ($S_-^2$) pair representing a composite local boson with kinematic restriction $(S_\pm^2)^2 = 0$, which underscores its hard-core boson nature. Local (on-site) nondiagonal order parameter $\langle S_\pm^2 \rangle$, which is effectively a parameter of local superconducting order, differs from zero only if a quantum superposition of states $|-1\rangle$ and $|+1\rangle$ is established at a site.

Introducing the pseudospin $S = 1$ formalism to characterize charge triplets, we write the effective Hamiltonian, which commutes with the $z$-component of total pseudospin $\Sigma_i S_{iz}$ and thus conserves the total charge of the system, as a sum of potential and kinetic energies:

$$H = H_{\text{pot}} + H_{\text{kin}}^{(2)}, \qquad (2)$$

where

$$H_{\text{pot}} = \sum_i (\Delta S_{iz}^2 - \mu S_{iz}) + \frac{1}{2} V \sum_{\langle ij \rangle} S_{iz} S_{jz}, \qquad (3)$$

and only the contribution of two-particle transfer of local composite bosons is taken into account in the kinetic energy:

$$H_{\text{kin}}^{(2)} = -\frac{1}{2} t_b \sum_{\langle ij \rangle} (S_{i+}^2 S_{j-}^2 + S_{i-}^2 S_{j+}^2). \qquad (4)$$

The first term in (3) (single-ion anisotropy) characterizes the on-site density–density correlation effects. Parameter $\Delta$ is related to the known correlation parameter $U$: $\Delta = U/2$. The second term may be associated with the pseudomagnetic field along axis $Oz$ or with the chemical potential with respect to the addition of new particles. The last term characterizes intersite interactions (correlations) of the density–density type. In subsequent analysis, only the interaction of nearest neighbors with positive (antiferromagnetic) parameters of intersite correlations $V$ is taken into account.

Depending on the ratio between the parameters of Hamiltonian (2) and on the total charge, the ground state of the system corresponds to a homogeneous nonconducting phase of the quantum paramagnetic type with $\langle S_z \rangle = \langle S_z^2 \rangle = 0$, which is established at large positive values of correlation parameter $\Delta$ (large-$U$ phase); to a nonconducting charge-order (CO) phase, which is equivalent to antiferromagnetic ordering along the $z$ axis; or to a superfluid (SF) phase with a nonzero order parameter $\langle S_\pm^2 \rangle$, which is accompanied by homogeneous ferromagnetic ordering or inhomogeneous antiferromagnetic ordering (supersolid phase) of $z$-components of the pseudospin. Local superconducting order parameter $\langle S_\pm^2 \rangle$ may be written in the standard form as $|\Psi| e^{\pm i\phi}$ with modulus $|\Psi|$ and phase $\phi$.

## 3. SPECIFIC FEATURES OF THE STRUCTURE OF ANTIPHASE DOMAIN BOUNDARIES OF THE CO PHASE

Using an NVIDIA graphics processing unit and the Monte Carlo method, we simulated the charge-ordering phase transition in the model cuprate in the two-sublattice approximation on a 256 × 256 square lattice with periodic boundary conditions and parameters $t_b = 1$, $V = 0.75$, and $\mu = 0$. This set of parameters ensures that a CO ground state is maintained in a sufficiently wide range of variation of local correlation parameter $\Delta$.

A branched domain structure formed in the process of rapid thermalization (annealing) at $\Delta = -5$. At low temperatures, well-marked filamentary superconductivity emerged at the center of antiphase domain boundaries of the CO phase, which is characterized by a nonzero modulus of the local superconducting order parameter. The latter fact is indicative of the presence of local quantum superpositions $Cu^{1+}$–$Cu^{3+}$. As the transfer integral of composite boson $t_b$ increases, the domain boundaries get wider, and the volume of the superconducting state increases to the point when the CO phase becomes suppressed completely and the transition to an inhomogeneous superconducting state occurs.

Interestingly, both the CO domain structure the superconducting structure of a domain boundary turned out to be stable with respect to large variations of local correlation parameter $\Delta$ and were still preserved at $\Delta \approx +1.0$. However, further enhancement of local correlations resulted in a fundamental rearrangement of the structure of domain boundaries. Figure 1 presents the pattern of evolution of an antiphase domain boundary at $\Delta \geq +1.0$, and Fig. 2 shows the phase distribution of the local superconducting order parameter (phase flow) in a domain boundary. When the value of $\Delta$ increases, the regular structure of filamentary superconductivity at the center of an antiphase domain boundary gets disrupted, and regions of the "parental" $Cu^{2+}$ phase (or, in pseudospin terms, the quantum paramagnetic phase) emerge. These regions grow with $\Delta$; at $\Delta \approx +1.4$, filamentary superconductivity becomes suppressed completely and the $Cu^{2+}$ phase occupies the entire boundary. At even higher $\Delta \geq +1.5$, the domain boundary widens, while charge ordering is suppressed gradually. In other words, the transition from the CO phase to the parental phase (large-$U$ phase) at higher values of the local correlation parameter is effected by the growth of domain boundaries.

The study of temperature effects demonstrates that transitions from the superconducting state to the parental phase and then to the disordered "paramagnetic" state occur in the CO phase domain boundaries as the temperature increases at $\Delta = +1.0$. However, subsequent cooling to very low temperatures $T = 0.0001$ results in the restoration of just the parental

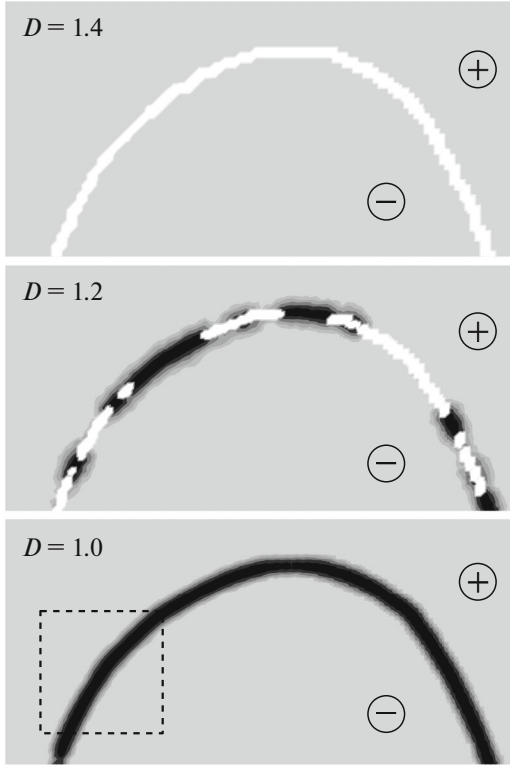

**Fig. 1.** Evolution of an antiphase domain boundary induced by the growth of local correlation parameter Δ. A fragment of the 256 × 256 lattice with an antiphase domain boundary separating CO domains denoted by the plus and minus signs is highlighted. Filamentary superconducting and parental $Cu^{2+}$ phases are colored black and white, respectively.

structure of domain boundaries; i.e., hysteretic behavior is observed.

## 4. CONCLUSIONS

The effect of the strength of local correlations $\Delta = U/2$ on the structure of domain boundaries of the CO phase of a model cuprate was studied. The results of numerical Monte Carlo modeling on large square lattices revealed the formation of a branched domain structure in the process of rapid annealing. Filamentary superconductivity, which remained stable in a wide range of $U$ variation up to $U \approx 2$, emerged in antiphase domain boundaries of this structure. However, as local correlations grew even stronger, filamentary superconductivity was disrupted, and the filamentary parental $Cu^{2+}$ phase, which separated domains with charge ordering $Cu^{1+}-Cu^{3+}$, formed in the boundaries. The modeling of temperature effects revealed hysteretic behavior of the boundary structure.

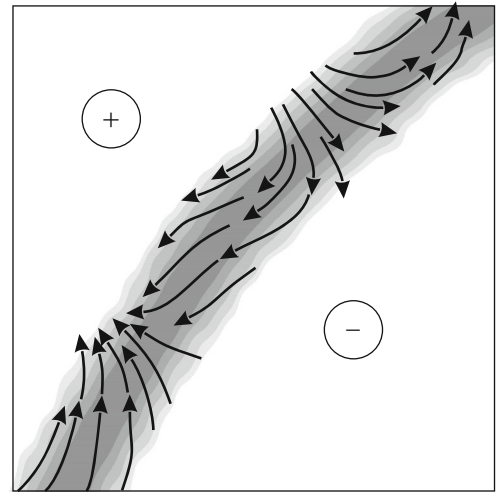

**Fig. 2.** Phase distribution of the local superconducting order parameter (phase flow) in the domain-boundary section marked in Fig. 1. Shades of gray highlight the inhomogeneous distribution of the modulus of the local superconducting order parameter.

## ACKNOWLEDGMENTS

This study was supported by Program 211 of the Government of the Russian Federation, agreement no. 02.A03.21.0006, and projects 2277 and 5719 of the Ministry of Education and Science of the Russian Federation.